\documentclass[12pt]{article}
\pdfoutput=1
\usepackage[english]{babel}
\usepackage{dirtytalk}
\usepackage[utf8x]{inputenc}
\usepackage{arxiv}

\usepackage{apacite}
\usepackage{filecontents}

\usepackage{amssymb,amsmath,amsfonts,eurosym,geometry,ulem,graphicx,caption,color,setspace,sectsty,comment,natbib, footmisc,caption,pdflscape,subfigure,array,hyperref}

\newcolumntype{L}[1]{>{\raggedright\let\newline\\arraybackslash\hspace{0pt}}m{#1}}
\newcolumntype{C}[1]{>{\centering\let\newline\\arraybackslash\hspace{0pt}}m{#1}}
\newcolumntype{R}[1]{>{\raggedleft\let\newline\\arraybackslash\hspace{0pt}}m{#1}}

\begin{document}

\begin{titlepage}
\title{Explaining Caste-based Digital Divide in India}
\author {R Vaidehi\thanks{Research Scholar, Department of Economics and  Finance, Birla Institute of Technology and Science-Pilani, Hyderabad campus, E-mail: p20170101@hyderabad.bits-pilani.ac.in}\and Bheemeshwar Reddy A\thanks{Assistant Professor, Department of Economics and  Finance, Birla Institute of Technology and Science-Pilani, Hyderabad campus, E-mail: bheem@hyderabad.bits-pilani.ac.in} \and  Sudatta Banerjee\thanks{Assistant Professor, Department of Economics and  Finance, Birla Institute of Technology and Science-Pilani, Hyderabad campus, E-mail: sudatta@hyderabad.bits-pilani.ac.in }}
\date{09-11-2020}
\maketitle
\begin{abstract}
 With the increasing importance of information and communication technologies in access to basic services like education and health, the question of digital divide based on caste assumes importance in India where large socioeconomic disparities persist between different caste groups. Studies on caste-based digital inequality are still scanty in India. Using nationally representative survey data, this paper analyzes the first-level digital divide (ownership of computer and access to the Internet) and the second-level digital divide (individual’s skill to use computer and the Internet) between the disadvantaged caste groups and others. Further, this paper identifies the caste group based differences in socioeconomic factors that contribute to the digital divide between these groups using a non-linear decomposition method. The results show that there exists large first-level and second-level digital divide between the disadvantaged caste groups and others in India.  The non-linear decomposition results indicate that caste-based digital divide in India is rooted in historical socioeconomic deprivation of disadvantaged caste groups. More than half of the caste- based digital gap is attributable to differences in educational attainment and income between the disadvantaged caste groups and others. Findings of this study highlight the urgent need for addressing  educational and income inequality between the different caste groups in India in order to bridge the digital divide.


\end{abstract}
\setcounter{page}{0}
\thispagestyle{empty}
\end{titlepage}

\section{Introduction} \label{sec:Introduction}
In the contemporary world, information and communication technologies (ICTs) are increasingly becoming crucial for citizens to actively take part in economic, political and social life \citep{roztocki2019}. Several researchers and policymakers argue that a wide adoption of ICTs enables economic development\citep{aker2010,donner2015,heeks2010,roztocki2019,saith2008,swaminathan2018}. Different international agencies continue to recommend the use of ICTs for achieving inclusive growth and for improving governance; for generating new economic opportunities for underserved populations; and, for the delivery of health, education and other public services across the world, especially so in developing countries \citep{world2016,atchoarena2017,kelly2018}. For example, the goal 9.c of Sustainable Development Goals (SDGs) of the United Nations is to \say{significantly increase access to ICT and strive to provide universal and affordable access to the Internet in the least developed countries by 2020}\citep{UNDP2018}. However, given the rising economic inequalities and persistence of poverty across the world, all individuals or social groups do not have equal access to ICTs. The inequality in access to ICTs is often referred to as the \say{Digital Divide}. The Organisation for Economic Co-operation and Development (OECD) defines the digital divide as \say{the gap between individuals, households, businesses and geographic areas at different socioeconomic levels with regard both to their opportunities to access information and communication technologies (ICTs) and to their use of the Internet for a wide variety of activities} \cite[p.~5]{oecd2001}.\\

Different studies have shown that inadequate access to digital technology can adversely affect the educational, health and labour market outcomes of the underprivileged social groups \citep{warren2007,mossberger2007,robinson2015,pagani2016, gao2018,haenssgen2018}. In the wake of the COVID-19 pandemic, educational institutions across the world had to shift to online teaching. However, as a result of the digital divide, a large number of poor students without computers and access to the Internet could not participate in online education \citep{van2020}, thus elucidating the role of digital divide in exacerbating the existing educational inequality between the digital-haves and digital-have nots. \cite{dimaggio2008} have shown that people who use the Internet earned higher wages than those who did not. \cite{peng2017} found that digital skills improve employment opportunities by lessening worker displacement and also enhancing the chances of reemployment for displaced workers. The digital divide can widen consumption inequality between households. In the context of increasing use of ICTs in providing health services in developing countries, studies have found that lack of access to digital technology among the marginalized groups can further increase health inequalities\citep{haenssgen2018}.\\ 

Only a limited number of studies have examined the digital divide in India. Especially, studies that have examined caste-based digital inequality using nationally representative data are sparse in India. The only exception to this is a recent study by \citep{tewathia2020}.  They have employed the network effects approach to examine the relationship between social and digital inequality in India. This study is based on the data from the Indian Human Development Survey 2011. They find that there are substantial disparities in ownership and usage of computer and the access to the Internet among different socioeconomic groups in India. They show that educational attainment, caste, and occupation of the household are some of the factors that are associated with ownership and use of computer and the Internet in India. Another study by \citeauthor{singh2013}(2013) based on a sample of 500 individuals across four states (Haryana, Madhya Pradesh, Punjab and Rajasthan) in the north and central parts of India found that education and some familiarity with the English language is a crucial determinant of computer and Internet use. Their results also show a significant difference between males and females in the use of computer and the Internet. \citeauthor{venkatesh2013} (2013,p.239) based on a longitudinal study of 210 families in rural India show that “the social network constructs contributed significantly to the explanation of technology use”. A recent study finds that there is a substantial regional divide in digitalization in India \citep{tewathia2020b}.\\

Caste is a unique marker of social stratification in South Asia, especially so in India. The caste system is characterized by graded socioeconomic inequality. Certain caste groups have historically faced social and economic discrimination, including in access to education in India. These disadvantaged caste groups include Scheduled Castes (SC), Scheduled Tribes (ST) and Other Backward Classes (OBC) which together account for around 80\% of the Indian population. To address the caste-based inequalities, the Indian state introduced reservations in education and employment for the disadvantaged caste groups as a constitutional guarantee since the early years of independence. However, several recent studies have shown that the disadvantaged caste groups continue to experience discrimination in labour market and caste-based inequalities in education, income and wealth persist in contemporary India \citep{borooah2005caste,borooah2005vidya,kijima2006caste,madheswaran2007caste,thorat2007legacy,deshpande2011grammar, zacharias2011caste, azam2012distributional}. Recent studies have also shown that the disadvantaged caste groups have lower chances of economic and educational mobility as compared to the non-disadvantaged in India \citep{motiram2012close,sharma2018measuring, asher2018intergenerational,choudhary2019inequality}.  \footnote{ Notably, the economic, educational and employment performance of Muslims, a religious minority in India, is found to be as worse as one of the most disadvantaged caste groups such as Scheduled Castes \citep {sachar2006social,desai2008changing,basant2010handbook, niaz2014social, kumar2020backward}.Taking cognisance of this fact, this paper additionally presents religion-based digital divide analysis in the Appendix, while the paper focuses exclusively on caste-based digital inequalities. In India, caste and religious categories often overlap. For the analysis in the Appendix, the Muslim social group includes all people who have reported their religion as Islam, irrespective of their caste. To ensure that Muslims are compared to the non-SC, ST and OBC population in other religions, the paper first divides the population into Muslims and the rest. Next, it separates the SC, ST and OBC from the non-Muslim population. For the analysis done in the Appendix, the Muslims are compared to this residual category which we refer to as the non-disadvantaged (non-SC, ST, OBC and Muslims) (See Appendix).}\\

Disparities in access to and use of ICTs can further deepen the existing socioeconomic inequalities between social groups in a multi-group society \citep{van2005deepening,helsper2012corresponding,hargittai2018digital,muschert2018theorizing,witte2010internet}. However, the research on digital inequality based on caste is scant in India. The existing studies cited above either deal with limited geographical areas, or having relied on data which is almost a decade old, do not throw light on the current levels of digital disparities. Further, owing to the lack of data on second-level digital divide (that is, digital skills) the existing studies engage only with the first-level digital divide (that is, the ownership of computers and access to the Internet) and consequently inequalities in digital skill remain unexplored.The present study uses the latest nationally representative sample survey data to analyse the caste-based digital divide. The significance of the present study also lies in explaining the caste-based digital divide in terms of the socioeconomic differences that exist between the caste groups. This exercise, to our knowledge, has not been undertaken in the context of India within the existing literature on digital divide. The present study examines the first and second-level digital inequality between Others and each of the disadvantaged caste  groups — SC, ST and OBC.\\ 

Rest of the paper is organized into following sections. The next section discusses the socioeconomic and demographic factors that are associated with the digital divide. The third and fourth sections discuss the data and methodology respectively. The fifth section presents results on how different factors contribute to the digital divide between the Other group and each of the disadvantaged caste groups in India. The final section presents discussion and conclusions.\\

\section{Factors associated with the digital divide} \label{sec:Factors associated with the digital divide}
The digital divide can be broadly categorized into three levels \citep{ragnedda2018three,scheerder2017determinants}. The first-level digital divide refers to the divide that is measured through access to ICTs (i.e., having a computer and access to the Internet) and is expressed as a binary outcome. The second-level digital divide refers to the gap in digital skills required to use computer and the Internet. The third-level digital divide measures \say{inequalities in the capacities to get the benefits from the access and use of the Internet} \citep[p.~6]{ragnedda2018three}. The first and second-level digital inequality at individual level, which is the focus of this paper, are mainly associated with factors such as differences in socioeconomic status as well as demographic factors including educational attainment, income level, occupation, social origin (race, caste, ethnicity, religion and  class), place of residence (urban or rural areas) and language \citep{compaine2001digital,mossberger2003virtual,fairlie2004race,helsper2010gendered,nishijima2017evolution}. In this section, we discuss some of the factors that explain the digital divide at individual level.\\

Differences in educational attainment is the most significant factor that is associated with the digital divide among individuals \citep{haight2014revisiting,van2011internet,vicente2010drives,cruz2016education}. Better educational attainment facilitates effective use of a computer, and also aids in absorption and comprehension of information received through the Internet \citep{hsieh2008understanding}. Hence, inequalities in educational attainment can lead to a digital divide between individuals with higher and lower educational attainment. Studies have also found that language can create a barrier for adopting digital technologies \citep{ono2008immigrants,singh2013, jara2015understanding}. Availability of content in the local language is critical to the use of the Internet.\\

Different studies have shown that income is a crucial predictor of the digital divide \citep{fairlie2004race,chaudhuri2005analysis,vicente2010drives}. Higher economic inequality with a high level of poverty may hinder the process of ICTs diffusion among poorer households. Individual’s occupation is also an essential factor associated with the digital divide. In other words, individuals employed in fields where application of ICTs are very common, such as scientific and technical workers, are more likely to use digital technologies than people employed in other occupations \citep{lengsfeld2011econometric,novo2014breaking}. Gender is a key factor that contributes to the digital divide \citep{bimber2000measuring,hargittai2006differences,martinez2017digital}. However, scholars differ on the extent of environmental and attitudinal factors in the differential use of ICTs by gender. \citeauthor{bimber2000measuring}(2000) shows that even after accounting for differences in income between males and females, women used the Internet less frequently than their male counterparts. On the contrary, \citeauthor{hilbert2011digital}(2011) emphasises the unfavourable conditions related to employment, education and income in the digital divide between men and women. Age is also significantly associated with the digital divide. Age is negatively related to the access to and use of ICTs. Across the world younger people are more likely to use computer and the Internet as compared to the older population \citep{loges2001exploring,abbey2009no}. Studies from different countries have shown that there exists a vast digital divide between rural and urban areas \citep{hindman2000rural,prieger2013broadband,salemink2017rural}. As compared to the population in rural areas, the urban population has better access to the Internet owing to better availability of supporting infrastructure such as electricity and broadband connectivity \citep{park2017digital, philip2017digital,salemink2017rural}. Geographical location of the individual, both at the global level as well as within countries has also been important in determining access to the digital world\citep{chinn2007determinants}. In India, the state-- a geographical and administrative region-- of residence of an individual, is also a major factor associated with the digital divide\citep{tewathia2020b}.\\

\section{Data} \label{sec:data}
To study the caste-based digital inequalities in India, this paper uses unit level data from survey on Household Social Consumption: Education, \citep{GOI2018}. This survey is a nationally representative multistage stratified household sample survey carried out by the National Sample Survey, Government of India between July 2017 and June 2018. The survey collected information on households’ access to and use of computer and the Internet in India. The survey defined a computer  as any \say{devices like, desktop computer, laptop computer, notebook, netbook, palmtop, tablet (or similar handheld devices)}\cite[p.~B-6-7]{GOI2019}. A household  is considered to be having access to Internet  \say{if a household member of age 5 years and above used Internet to find, evaluate and communicate information from any location during the last 30 days preceding the date of survey, via any device, like, desktop, laptop, palmtop, notebook, netbook, smartphone, tablets, etc., }\cite[p.~B-7]{GOI2019}. In the context of the rising use of budget smartphones in developing countries,  it is important to account for the use of smartphones as an alternative to computers. While the definition of 'computer' does not include smartphones, the use of smartphones to access the Internet is captured in the definition of household members' access to the  Internet in the survey. This allows for comprehensive analysis of digital divide, representing both computer and smartphone usage in India.\\

        The data set contains detailed information on various socioeconomic and demographic characteristics of each member of the household. The survey provides data on whether the household has a computer and has access to the Internet. Further, the survey has recorded information on each member (five years and above) of the household regarding their ability to operate the computer and use the Internet. The use of the Internet in the last 30 days was also ascertained from each individual of the sample households.\footnote{The survey considered a person knowledgeable in the use of computers, if he/she was capable of carrying out any of the specified tasks using a computer. These tasks included, \say{copying or moving a file or folder, using copy and paste tools to duplicate or move information within a document, sending e-mails with attached files (e.g. document, picture, and video), using basic arithmetic formulae in a spreadsheet, connecting and installing new devices (e.g. modem, camera, printer), finding, downloading, installing and configuring software, creating electronic presentations with presentation software (including text, images, sound, video or charts), transferring files between a computer and other devices, writing a computer program using a specialized programming language.} \cite[p.~B-7]{GOI2019}. The survey considered a person knew how to use the Internet, if he/she could perform any of the following activities using the Internet on any device — \say{use Internet browser for website navigation, using e-mail and social networking applications, etc., to find, evaluate and communicate information} \cite[p.~B-8]{GOI2019}}\\

The present paper uses this data to analyze the digital divide at individual level between the  Other group and each of the disadvantaged caste groups in India. As discussed earlier, the disadvantaged caste groups include STs, SCs and OBCs. Consequently, the Other  group consists of the rest of the Indian population and is a residual category.\\

To examine the digital divide between Other and each of the disadvantaged caste groups, the computer ownership rate,  the Internet access rate, the computer literacy rate, the Internet literacy rate and the Internet use rate are considered as outcome variables. The computer ownership rate for a caste group is simply the proportion of individuals having computer at home. Similarly, the Internet access rate for a caste group represents the share of individuals having Internet facility at home. The computer literacy rate for a caste group is defined as the proportion of individuals from the caste group who know how to operate a computer. The share of individuals in a caste group knowing how to use the Internet is the Internet literacy rate for the caste group. The Internet use rate for a caste group is defined as the proportion of individuals in that particular caste group who have used the Internet during the last 30 days from the date when the survey was conducted.\\

We restrict our analysis to individuals in the age group between 15 years and 59 years, which has resulted in a sample of 349,914 observations. In the present study, all outcome variables are binary variables. For analysis of the digital divide, the covariates include individual's age, sex, number of years of education, occupation (type of employment) of the head of household and per capita income of the household, place of residence (rural or urban) and state (geographical and administrative unit in India).  Monthly per capita expenditure of households is considered as a proxy for per capita income. The occupation or type of employment  of the head of the household, to which an individual belongs, is classified into six categories: self-employed in agriculture, self-employed in non-farm, regular worker, casual works in non-agriculture, casual worker in agriculture and other workers. Digital infrastructure in India varies substantially between rural and urban areas as well as across the states in India. The state variable in the analysis represents the state in which the individual is residing. 

\section{Methodology} \label{sec:Methodology}
The Oaxaca-Blinder decomposition (Blinder, 1973; Oaxaca, 1973), a linear regression decomposition approach, is used to explain the contribution of factors that drive average socioeconomic outcomes such as wages or income gap between two groups. The Oaxaca-Blinder method allows to decompose the average gap in an outcome variable, $Y$, between two groups $a$ and $d$ into the following :
$$ \bar{Y^{a}}-\bar{Y^{d}} =\left[ \left(\bar{X_{a}}-\bar{X_{d}}\right)\hat{\beta_{a}} \right]+ \left[\bar{X_{d}} \left(\hat{\beta_{a}}-\hat{\beta_{d}}\right) \right] \hspace{2cm}\left(1\right)  $$
where $X_{i}$  represents the vector of average values of the covariates and $\hat{\beta_{i}}$  is a vector of coefficients for group $i$. The Oaxaca-Blinder decomposition separates the mean difference in an outcome between two groups into two parts \say{explained} and \say{unexplained} differences \citep{blinder1973wage,oaxaca1973male}. In this study, we focus on the \say{explained} part as this can inform policy that seek to bridge the digital gap between different caste groups in India.\\

Given that the outcome variables such as possession of computer or the use of the Internet by individuals is binary in this study, a direct application of Oaxaca-Blinder linear decomposition \citep{blinder1973wage,oaxaca1973male} is not possible. Hence, this paper employs Fairlie (2005) non-linear decomposition method. The Fairlie (2005) non-linear decomposition is an adaptation of Oaxaca-Blinder decomposition for binary dependent variables. Earlier studies have used the Fairlie non-linear decomposition method for analyzing the digital divide based on race and ethnicity \citep{fairlie2004race,fairlie2017have,manlove2019understanding}. Following Fairlie (2005), the decomposition for a non-linear model of the type  $P(Y=1)=F(X\hat{\beta})$ for the mean digital gap between Others ($a$) and a disadvantaged caste group ($d$) can be divided into following components:\\
$$  \bar{Y^{a}}-\bar{Y^{d}}= \left[ \sum\limits_{\forall{N^{a}}} \frac{F(X_{i}^{a}\hat{\beta}^{a})}{N^{a}}- \sum\limits_{\forall{N^{d}}} \frac{F(X_{i}^{d}\hat{\beta}^{a})}{N^{d}}\right]+ \left[ \sum\limits_{\forall{N^{d}}} \frac{F(X_{i}^{d}\hat{\beta}^{a})}{N^{d}}- \sum\limits_{\forall{N^{d}}} \frac{F(X_{i}^{d}\hat{\beta}^{d})}{N^{d}}\right] \hspace{2cm}\left(2\right) $$\\
The first component of the equation (2) represents the share of the digital gap that is attributed to differences in the distribution of covariates between Others and a disadvantaged caste group. It is often referred to as the \say{explained part} or the \say{covariate effect}. The other component is called the \say{unexplained part} or the \say{coefficient effect}. The unexplained part of the digital gap arises due to differences in the effect of covariates between two caste groups.\\

Alternatively, the non-linear decomposition of the mean digital gap between two caste groups can also be written as:\\ 
$$  \bar{Y^{a}}-\bar{Y^{d}}= \left[ \sum\limits_{\forall{N^{a}}} \frac{F(X_{i}^{a}\hat{\beta}^{d})}{N^{a}}- \sum\limits_{\forall{N^{d}}} \frac{F(X_{i}^{d}\hat{\beta}^{d})}{N^{d}}\right]+ \left[ \sum\limits_{\forall{N^{d}}} \frac{F(X_{i}^{a}\hat{\beta}^{a})}{N^{d}}- \sum\limits_{\forall{N^{a}}} \frac{F(X_{i}^{a}\hat{\beta}^{d})}{N^{a}}\right] \hspace{2cm}\left(3\right) $$
The equation (2) and (3) differ in terms of weights used. As against equation (2), the first component of the equation (3) is weighted by a disadvantaged caste group’s regression coefficients($\hat{\beta}^{d}$) and the second component is weighted by the Other’s average covariates ($X^a$). Equation (2) and (3) intuitively have the same meaning, though they may yield different results. Following the recommendations of \citeauthor{oaxaca1994discrimination}(1994), the study has  carried out decomposition analysis using coefficient estimates from a pooled sample to weight the explained part of the decomposition. 

Equation (4) identifies the individual contribution of each of the covariates $X_i$.\\
$$\frac{1}{N^{d}}\sum\limits_{\forall{N^{d}}} F\left[  \hat{\alpha^* }+X_{1i}^{a} \hat{\beta}_{1}^{*}+X_{2i}^{a} \hat{\beta}_{2}^{*}\right]-F\left[  \hat{\alpha^* }+X_{1i}^{d} \hat{\beta}_{1}^{*}+X_{2i}^{a} \hat{\beta}_{2}^{*}\right] \hspace{2cm}\left(4\right)$$
In equation (4), $\hat{\beta^{*}}$ denotes probit coefficients from the pooled sample. Here, the sample sizes of two caste groups are not comparable. To deal with this issue, following \citep{fairlie2005extension} the paper selects a random sub-sample from the larger caste group equal to the size of the smaller caste group and averages the decomposition results of 1000 sub-samples. Finally, as results are sensitive to the ordering of the independent variables, the order for each iteration is randomised to address the concern of path-dependency of results.

\section{Results}{\label{sec:Results}}
\subsection{Descriptive statistics }

\begin{table}[htbp]
  \centering
  \caption{Summary statistics}
   \resizebox{\textwidth}{!}{ \begin{tabular}{llllllll}
\hline
      & Others(1) & ST(2) & SC(3) & OBC(4) & (1)-(2) & (1)-(3) & (1)-(4) \\
      \hline
    Variables & Mean & Mean & Mean & Mean & t-test & t-test & t-test \\
    \hline
    COR & 0.207 & 0.055 & 0.066 & 0.095 & 0.153*** & 0.142*** & 0.113*** \\
      & [0.001] & [0.001] & [0.001] & [0.001] &   &   &  \\
    IAR & 0.411 & 0.141 & 0.182 & 0.243 & 0.270*** & 0.229*** & 0.168*** \\
      & [0.002] & [0.002] & [0.002] & [0.001] &   &   &  \\
    CLR & 0.312 & 0.112 & 0.135 & 0.189 & 0.200*** & 0.176*** & 0.123*** \\
      & [0.001] & [0.001] & [0.001] & [0.001] &   &   &  \\
    ILR & 0.380 & 0.145 & 0.176 & 0.238 & 0.235*** & 0.204*** & 0.143*** \\
      & [0.001] & [0.002] & [0.002] & [0.001] &   &   &  \\
    IUR & 0.347 & 0.123 & 0.150 & 0.206 & 0.224*** & 0.197*** & 0.141*** \\
      & [0.001] & [0.001] & [0.001] & [0.001] &   &   &  \\
    Age & 34.625 & 33.489 & 33.531 & 34.006 & 1.137*** & 1.095*** & 0.619*** \\
      & [0.038] & [0.055] & [0.051] & [0.032] &   &   &  \\
    Urban & 0.444 & 0.124 & 0.226 & 0.300 & 0.320*** & 0.218*** & 0.145*** \\
      & [0.002] & [0.001] & [0.002] & [0.001] &   &   &  \\
    Male & 0.515 & 0.507 & 0.516 & 0.509 & 0.008*** & -0.001 & 0.007*** \\
      & [0.002] & [0.002] & [0.002] & [0.001] &   &   &  \\
    Log(Income) & 7.793 & 7.256 & 7.396 & 7.508 & 0.537*** & 0.397*** & 0.285*** \\
      & [0.002] & [0.002] & [0.002] & [0.001] &   &   &  \\
    Education & 9.638 & 6.384 & 7.025 & 7.868 & 3.254*** & 2.612*** & 1.770*** \\
      & [0.015] & [0.023] & [0.022] & [0.014] &   &   &  \\
    SEA & 0.272 & 0.395 & 0.225 & 0.325 & -0.123*** & 0.047*** & -0.053*** \\
      & [0.001] & [0.002] & [0.002] & [0.001] &   &   &  \\
    SENA & 0.290 & 0.121 & 0.162 & 0.239 & 0.169*** & 0.128*** & 0.051*** \\
      & [0.001] & [0.001] & [0.002] & [0.001] &   &   &  \\
    RSW & 0.249 & 0.095 & 0.155 & 0.161 & 0.154*** & 0.094*** & 0.088*** \\
      & [0.001] & [0.001] & [0.002] & [0.001] &   &   &  \\
    CWA & 0.058 & 0.181 & 0.197 & 0.093 & -0.123*** & -0.140*** & -0.036*** \\
      & [0.001] & [0.002] & [0.002] & [0.001] &   &   &  \\
    CWNA & 0.076 & 0.165 & 0.220 & 0.137 & -0.089*** & -0.144*** & -0.062*** \\
      & [0.001] & [0.002] & [0.002] & [0.001] &   &   &  \\
    OW & 0.035 & 0.024 & 0.019 & 0.026 & 0.011*** & 0.016*** & 0.010*** \\
      & [0.001] & [0.001] & [0.001] & [0.000] &   &   &  \\
    Observations & 106540 & 49420 & 57848 & 142062 &   &   &  \\
\hline
    \multicolumn{8}{p{40.15em}}{Notes: Computer ownership rate (COR); Internet access rate (IAR); Computer literacy rate (CLR); Internet literacy rate (ILR); Internet use rate (IUR);  Self-employed in agriculture (SEA); Self-employed in non-agriculture (SENA); Regular salaried Workers (RSW); Casual worker in agriculture (CWA); Casual worker in non-agriculture (CWNA); Other work (OW). Standard errors are in parentheses. ***, **, and * indicate significance at the 1, 5, and 10 percent critical level. Survey sampling weights are used for the estimation.} \\
    \end{tabular}}%
  \label{tab:addlabel}%
\end{table}%

 Table 1 presents mean and corresponding standard errors for each outcome variable and independent variable by caste group and t-test results for gap in outcome and independent variables between the Others and each of the disadvantaged caste groups. All the disadvantaged caste groups fare poorly with respect to every outcome variable--- computer ownership rate, Internet access rate, computer literacy rate, Internet literacy rate and Internet use rate. Only 14.1\% of the STs have access to the Internet as compared to 41.1\% of individuals from the  Other  group. Differences in the computer literacy rate across the caste groups are also pronounced, with only 11.2\% of STs and 13.5\% of SCs knowing how to use a computer. The same statistic for OBCs  is 18.9\% . However, the corresponding figure for the Others  is 31.2\%. Compared to STs, SCs, and OBCs , the Others are more affluent and better educated. For example, mean years of schooling for Others is 9.3, whereas the same for STs and SCs is just around 7 years; and, it is 7.8 years for OBCs. A much higher proportion of Others live in urban areas as compared to SC and ST individuals. For instance, 45\% of Other individuals live in urban areas with better access to digital infrastructure. The corresponding percentage for ST individuals is only 12.4\%. The lower socioeconomic position of the disadvantaged caste groups is also reflected in their occupational distribution. For example, 19.7\% of SC individuals belong to the category of 'casual workers in agriculture households' which is considered as a last resort in terms of occupation because of the poor working conditions and low wages. On the contrary, the corresponding share for the Others is only 5.8\%. Clearly, the overall socioeconomic status of the Otherss is significantly better than the rest of the caste groups.\\ 

\begin{table}[htbp]

  \centering
  \caption{Marginal effects from probit regressions on the probability of having computer at home, internet access at home, computer literacy, Internet literacy and Internet use. }
   \resizebox{\textwidth}{!}{\begin{tabular}{llllll}
\hline
      
   Variables& 	CO &	AIF & 	CL& IL &IU \\\\
   \hline
      Age & 0.001*** & 0.002*** & -0.003*** & -0.006*** & -0.005*** \\
    Gender: reference = Male \\
    Female & 0.012*** & 0.027*** & -0.022*** & -0.069*** & -0.067*** \\
      & (0.001) & (0.003) & (0.001) & (0.002) & (0.002) \\
     Place of residence: reference  = Rural &   &   &   &   &  \\
    Urban & 0.022*** & 0.055*** & 0.016*** & 0.034*** & 0.031*** \\
      & (0.002) & (0.004) & (0.001) & (0.003) & (0.002) \\
    Caste group: reference = Other &   &   &   &   &  \\
    ST & -0.029*** & -0.100*** & -0.019*** & -0.046*** & -0.039*** \\
      & (0.003) & (0.005) & (0.002) & (0.003) & (0.003) \\
    SC & -0.035*** & -0.071*** & -0.018*** & -0.042*** & -0.038*** \\
      & (0.002) & (0.005) & (0.002) & (0.003) & (0.003) \\
    OBC & -0.035*** & -0.055*** & -0.013*** & -0.028*** & -0.025*** \\
      & (0.002) & (0.004) & (0.001) & (0.002) & (0.002) \\
    Education & 0.008*** & 0.020*** & 0.017*** & 0.032*** & 0.026*** \\
      & (0.000) & (0.000) & (0.000) & (0.000) & (0.000) \\
    Log(Income) & 0.072*** & 0.127*** & 0.032*** & 0.069*** & 0.065*** \\
      & (0.002) & (0.003) & (0.002) & (0.002) & (0.002) \\
    Occupation : reference = SEA &   &   &   &   &  \\
    SENA  & 0.039*** & 0.065*** & 0.009*** & 0.028*** & 0.031*** \\
      & (0.003) & (0.005) & (0.001) & (0.003) & (0.003) \\
     CWA & -0.026*** & -0.073*** & -0.006*** & -0.018*** & -0.017*** \\
      & (0.002) & (0.005) & (0.001) & (0.003) & (0.002) \\
    CWNA  & -0.026*** & -0.060*** & -0.009*** & -0.009*** & -0.007** \\
      & (0.002) & (0.005) & (0.001) & (0.003) & (0.002) \\
    RSW & 0.051*** & 0.093*** & 0.027*** & 0.055*** & 0.057*** \\
      & (0.003) & (0.005) & (0.002) & (0.003) & (0.003) \\
      OW & 0.024*** & 0.025** & 0.034*** & 0.057*** & 0.046*** \\
      & (0.005) & (0.009) & (0.004) & (0.007) & (0.006) \\
  Observations & 349914 & 349914 & 349913 & 349913 & 349913 \\
    \hline
    \multicolumn{6}{p{40.15em}}{Notes: Computer ownership (CO); Access to Internet facility (AIF);  Computer literacy (CL); Internet literacy  (IL); Internet use  (IU); Self-employed in agriculture (SEA); Self-employed in non-agriculture (SENA);Regular salaried workers (RSW); Casual worker in agriculture (CWA);  Casual worker in non-agriculture (CWNA); Other work (OW). Robust standard errors are in parentheses. ***, **, and * indicate significance at the 1, 5, and 10 percent critical level. Survey sampling weights are used for the estimation.}\\

    \end{tabular}}%
  \label{tab:addlabel}%
\end{table}%

Table 2 reports  marginal effects from probit regressions on the probability of individuals having a computer at home, Internet access, computer literacy, Internet literacy and Internet use. The estimates indicate that there are substantial gaps in outcome variables between our reference category, Others, and each of the disadvantaged caste groups (SCs, STs, and OBCs). In other words, our probit analysis confirms that caste-based digital divide, measured using different indicators, persists despite controlling for other socioeconomic and demographic variables in our model. A recent study also finds that caste is an important predictor of access to ICTs in India \citep{tewathia2020}.  As shown in the earlier studies \citep{fairlie2004race}, socioeconomic status of individuals, such as income, positively correlates with having a computer and use of Internet. Educational attainment  of individuals is one of the important factors that is associated with the individual's access to Internet and computer literacy. This finding is in tandem with the existing studies \citep{haight2014revisiting,cruz2016education}. Further,  concurring with various studies from different countries\citep{bimber2000measuring,hargittai2006differences,singh2013, martinez2017digital} the estimates indicate that there is a significant gender based digital gap between males and females. Females are less likely to use the Internet and be computer literate as compared to men. Living in urban areas is associated with better digital outcomes. In other words, there is a considerable rural-urban digital gap in India.\\  

\begin{table}[htbp]
  \centering
  \caption{Nonlinear decompositions of caste group disparities in computer ownership rate}
     \resizebox{\textwidth}{!}{\begin{tabular}{lllllll}
     \hline 
         Variables & Others-ST & \% explained & Others-SC & \% explained & Others-OBC & \% explained   \\
              \hline 
    Gap in COR & 15.3 &   & 14.2 &   & 11.3 &  \\
    Age & 0.003*** & 2.0 & 0.003*** & 21.0 & 0.002*** & 1.8 \\
      & (0.000) &   & (0.000) &   & (0.000) &  \\
    Female & 0.000* & 0.0 & 0.000*** & 0.0 & 0.000 & 0 .0\\
      & (0.000) &   & (0.000) &   & (0.000) &  \\
    Education & 0.035*** & 22.9 & 0.032*** & 22.5 & 0.022*** & 19.5 \\
      & (0.001) &   & (0.001) &   & (0.001) &  \\
    Log(income) & 0.070*** & 45.8 & 0.063*** & 44.4 & 0.044*** & 39.0 \\
      & (0.002) &   & (0.002) &   & (0.001) &  \\
    Place of residence  & 0.027*** & 17.7 & 0.017*** & 12.0 & 0.010*** & 8.9 \\
      & (0.002) &   & (0.001) &   & (0.001) &  \\
    Occupation & 0.009*** & 5.9 & 0.010*** & 7.0 & 0.008*** & 7.1 \\
      & (0.001) &   & (0.001) &   & (0.000) &  \\
    State & -0.005** & -3.3 & -0.019*** & -13.4 & -0.016*** & -14.2 \\
      & (0.002) &   & (0.001) &   & (0.001) &  \\
    Total explained  & 155958 & 91.0 & 164383 & 74.7 & 248598 & 62.1 \\
    \hline 
    \multicolumn{7}{p{40.58em}}{Notes: Computer ownership rate(COR). Bootstrap standard errors are in parentheses(1000 replications).  ***, **, and *indicate significance at the 1, 5, and 10 percent critical level. Survey sampling weights are used for the estimation.}
    \end{tabular}}%
  \label{tab:addlabel}%
\end{table}%
The nonlinear decomposition approach has been employed to examine the contribution of caste-based differences in socioeconomic variables to the digital gap between Others and each of the disadvantaged caste groups. Table 3 reports decomposition estimates of the gap in computer ownership rate. The first row reports the gap in computer ownership rate between Others and each of the disadvantaged caste groups. The second row from the bottom presents the digital gap explained by all covariates in the model between Others and STs, SCs and OBCs  separately. \\

The gap in computer ownership rate between the Others and STs is 15.3 percentage points. The differences in covariates explain 91\% ((0.139/0.153)*100) of the gap in household computer ownership rate between Others and STs (see Column 2, Table 3). All the estimates of decomposition of gap between the Others and STs are positive except state. The positive coefficients imply that if STs had the same characteristics as the Others, their computer ownership rate would be higher. Differences in per capita income between Others and STs  has contributed to 45.8\% of the total gap (15.3) in computer ownership rate between them. Similarly, differences in the educational attainment between Others and STs accounted for 22.9\% of the total gap in computer ownership rate between them. Other characteristics such as occupation, place of residence and age together explained around one-fourth of the total gap in computer ownership rate between the Others and STs.\\
\begin{table}[htbp]
  \centering
  \caption{Nonlinear decompositions of caste group disparities in Internet access rate}
\resizebox{\textwidth}{!}{\begin{tabular}{lllllll}
     \hline 
        Variables & Others-ST & \% explained & Others-SC & \% explained & Others-OBC & \% explained   \\
              \hline 
   Gap in  IAR & 27.0 &   & 22.9 &   & 16.8 &  \\
    Age & 0.003*** & 1.1 & 0.003*** & 0.1 & 0.002*** & 0.1\\
      & (0.000) &   & (0.000) &   & (0.000) &  \\
    Gender & -0.000 & 0.0 & 0.000** & 0.0 & -0.000 & 0 .0\\
      & (0.000) &   & (0.000) &   & (0.000) &  \\
    Education & 0.063*** & 23.4 & 0.052*** & 22.7 & 0.036*** & 21.4 \\
      & (0.002) &   & (0.001) &   & (0.001) &  \\
    Log(income)  & 0.083*** & 30.8 & 0.059*** & 25.8 & 0.045*** & 26.8 \\
      & (0.003) &   & (0.002) &   & (0.001) &  \\
     Place of residence   & 0.039*** & 14.5 & 0.027*** & 11.8 & 0.014*** & 8.3 \\
      & (0.002) &   & (0.001) &   & (0.001) &  \\
    Occupation & 0.008*** & 3.0 & 0.015*** & 6.6 & 0.008*** & 4.8 \\
      & (0.001) &   & (0.001) &   & (0.000) &  \\
    State & 0.008*** & 3.0 & 0.010*** & 4.4 & 0.010*** & 6.0 \\
      & (0.002) &   & (0.002) &   & (0.002) &  \\
    Total explained &   & 75.7 &   & 72.6 &   & 68.4 \\
    \hline 
    \multicolumn{7}{p{40.58em}}{Notes: Internet access  rate(IAR).   ***, **, and *indicate significance at the 1, 5, and 10 percent critical level. Survey sampling weights are used for the estimation}
    \end{tabular}}%
  \label{tab:addlabel}%
\end{table}%

Similar to the STs, in the case of SCs and OBCs too, income and education emerge as the most significant contributors in explaining the gap in computer ownership rate between them and Others.  The negative coefficients of the state indicate that differences in state characteristics between them and the Others are in favor of STs, SCs and OBCs . For instance, if SC individuals had the same distribution across Indian states as the Other individuals, the computer ownership gap rate between them would have been 10.5\% higher. Our decomposition results are similar to Fairlie (2004) who found that income and education are the most important factors that explain digital gap between white and black racial groups in the USA.\\

\begin{table}[htbp]
  \centering
  \caption{Nonlinear decompositions of caste group disparities in computer literacy rate}
      \resizebox{\textwidth}{!}{\begin{tabular}{lrrrrrr}
      \hline
    Variables & \multicolumn{1}{l}{CLR} & \multicolumn{1}{l}{\% explained} & \multicolumn{1}{l}{CLR} & \multicolumn{1}{l}{\% explained} & \multicolumn{1}{l}{Gap in CLR} & \multicolumn{1}{l}{\% explained} \\
     \hline
    CLR & 20.0 &   & 17.7 &   & 12.3 &  \\
    Age & \multicolumn{1}{l}{-0.014***} & -7.0 & \multicolumn{1}{l}{-0.014***} & -7.9 & \multicolumn{1}{l}{-0.009***} & -7.3 \\
      & \multicolumn{1}{l}{(0.000)} &   & \multicolumn{1}{l}{(0.000)} &   & \multicolumn{1}{l}{(0.000)} &  \\
    Gender & \multicolumn{1}{l}{-0.000} & 0.0 & \multicolumn{1}{l}{-0.001***} & -0.6 & \multicolumn{1}{l}{-0.000***} & 0.0 \\
      & \multicolumn{1}{l}{(0.000)} &   & \multicolumn{1}{l}{(0.000)} &   & \multicolumn{1}{l}{(0.000)} &  \\
    Education & \multicolumn{1}{l}{0.093***} & 46.5 & \multicolumn{1}{l}{0.074***} & 41.9 & \multicolumn{1}{l}{0.053***} & 43.1 \\
      & \multicolumn{1}{l}{(0.002)} &   & \multicolumn{1}{l}{(0.001)} &   & \multicolumn{1}{l}{(0.001)} &  \\
    Log(income) & \multicolumn{1}{l}{0.063***} & 31.5 & \multicolumn{1}{l}{0.050***} & 28.3 & \multicolumn{1}{l}{0.036***} & 29.3 \\
      & \multicolumn{1}{l}{(0.002)} &   & \multicolumn{1}{l}{(0.002)} &   & \multicolumn{1}{l}{(0.001)} &  \\
    Place of residence  & \multicolumn{1}{l}{0.024***} & 12.0 & \multicolumn{1}{l}{0.018***} & 10.2 & \multicolumn{1}{l}{0.011***} & 8.9 \\
      & \multicolumn{1}{l}{(0.002)} &   & \multicolumn{1}{l}{(0.001)} &   & \multicolumn{1}{l}{(0.001)} &  \\
    Occupation & \multicolumn{1}{l}{0.011***} & 5.5 & \multicolumn{1}{l}{0.011***} & 6.2 & \multicolumn{1}{l}{0.008***} & 6.5 \\
      & \multicolumn{1}{l}{(0.001)} &   & \multicolumn{1}{l}{(0.001)} &   & \multicolumn{1}{l}{(0.000)} &  \\
    State & \multicolumn{1}{l}{-0.002} & -1.0 & \multicolumn{1}{l}{-0.001} & -0.6 & \multicolumn{1}{l}{-0.005***} & -4.1 \\
      & \multicolumn{1}{l}{(0.002)} &   & \multicolumn{1}{l}{(0.001)} &   & \multicolumn{1}{l}{(0.001)} &  \\
    Total explained (\%) &   & 87.5 &   & 77.6 &   & 76.5 \\
     \hline
    \multicolumn{7}{p{40.58em}}{Notes: Computer literacy rate(CLR).   ***, **, and *indicate significance at the 1, 5, and 10 percent critical level. Survey sampling weights are used for the estimation.}
    \end{tabular}}%
  \label{tab:addlabel}%
\end{table}%

Table 4 reports estimates of decomposition of the gap in the Internet access rate between the Others and each of the disadvantaged caste groups. The gap in the Internet access rate between the Others and ST individuals was the highest amongst all disadvantaged caste groups, i.e., ST individuals have 27 percentage points lower access to the Internet as compared to the Other individuals. The covariate effect, that is, the explained part of the total gap in the Internet access rate between the Other and ST individuals is 75.7\%. Among all the explanatory variables, the difference in individual’s education and per capita income are the most crucial factor that explain the gap in the Internet access rate between the Others and each of the disadvantaged caste groups. The individual's educational attainment alone contributes around one-fourth of the total gap in the Internet access rate between the Others and ST, SC and OBC individuals. After education and per capita income, disparities in place of residence contributes the most to the gap in the Internet access rate between the Others and each of the disadvantaged caste groups.  The differences in the distribution of population in urban areas contributes  14.5\% of the gap between the Others and STs. The corresponding figure for OBCs is only 8.3\%. As shown in Table 1, unlike STs a much higher proportion of OBCs live in the urban areas in India. \\

Table 5 presents the results of the decomposition of the computer literacy rate gap between Others and each of the disadvantaged caste groups. The gap in computer literacy rate between the Others and each of the disadvantaged caste groups is positive. The decomposition estimates of Others and STs and Others and OBCs show that differences in covariates explain around three fourth of the total gap in the computer literacy rate.  The differences in covariates between  Others and STs  account for 87.5\% of the total gap between their computer literacy rate. \\

The differences in the educational attainments of individuals from Others and each of the disadvantaged caste groups is the most critical factor that explains gaps  in computer literacy rate between them. For example, nearly half of the total gap in the computer literacy rate between Others and ST individuals is explained by differences in their educational attainment. The contribution of income is important in explaining the gap in the computer literacy rate between Others and each of the disadvantaged caste groups. Per capita income explains more than one-fourth of the total gap between Other and SC individuals. If OBCs had the same income as the Others, the gap in computer literacy rate between the Others and OBC individuals would decrease by 29.3\%. Other covariates, like the occupation of the head of household and individual’s place of residence, also contribute to the gap between the Others and the rest of the caste  groups. The state to which an individual belongs explains a tiny proportion of the gap. The negative sign of age shows that the gap between Others and STs in computer literacy rate would be 7 percentage points higher if STs had a similar age distribution as the Others.\\

\begin{table}[htbp]
  \centering
  \caption{Nonlinear decompositions of caste group disparities in Internet literacy rate}
    \resizebox{\textwidth}{!}{\begin{tabular}{lllllll}
     \hline 
        Variables & Others-ST & \% explained & Others-SC & \% explained & Others-OBC & \% explained   \\
              \hline 
    CLR & 0.235 &   & 0.204 &   & 0.143 &  \\
    Age & -0.015*** & -6.4 & -0.014*** & -6.9 & -0.008*** & -5.6 \\
      & (0.000) &   & (0.000) &   & (0.000) &  \\
    Gender & 0.000 & 0.0 & -0.000*** & 0.0 & -0.000 & 0.0 \\
      & (0.000) &   & (0.000) &   & (0.000) &  \\
    Education & 0.105*** & 44.6 & 0.085*** & 41.7 & 0.059*** & 41.4 \\
      & (0.002) &   & (0.001) &   & (0.001) &  \\
    Log(Income) & 0.070*** & 29.7 & 0.052*** & 25.5 & 0.036*** & 25.3 \\
      & (0.002) &   & (0.002) &   & (0.001) &  \\
    Place of residence  & 0.026*** & 11.0 & 0.019*** & 9.3 & 0.011*** & 7.7 \\
      & (0.002) &   & (0.001) &   & (0.001) &  \\
    Occupation & 0.010*** & 4.2 & 0.009*** & 4.4 & 0.006*** & 4.2 \\
      & (0.001) &   & (0.001) &   & (0.000) &  \\
    State & -0.001 & -0.4 & 0.006*** & 2.9 & 0.002 & 1.4 \\
      & (0.002) &   & (0.001) &   & (0.001) &  \\
    Total explained (\%) & 155957 & 82.9 & 164383 & 77.1 & 248598 & 74.4 \\
 \hline
    \multicolumn{7}{p{40.58em}}{Notes: Internet literacy rate (ILR).  ***, **, and * indicate significance at the 1, 5, and 10 percent critical level. Survey sampling weights are for used the estimation.}
    \end{tabular}}%
  \label{tab:addlabel}%
\end{table}%

\begin{table}[htbp]
  \centering
  \caption{Nonlinear decompositions of caste group disparities in Internet use rate}
    \resizebox{\textwidth}{!}{\begin{tabular}{lrrrrrr}
     \hline 
                    Variables  & \multicolumn{1}{l}{ Others-ST} & \% explained  & \multicolumn{1}{l}{Others-SC} &\% explained& \multicolumn{1}{l}{ Others-OBC} &  \% explained\\

              \hline 
                  CLR & \multicolumn{1}{l}{22.4} &  & \multicolumn{1}{l}{19.7} & -7.6 & \multicolumn{1}{l}{14.1} &  \\

Age & \multicolumn{1}{l}{-0.014***} & -6.2 & \multicolumn{1}{l}{-0.015***} & -7.6 & \multicolumn{1}{l}{-0.009***} & -6.4 \\
   Gender & \multicolumn{1}{l}{(0.000)} &   & \multicolumn{1}{l}{(0.000)} &   & \multicolumn{1}{l}{(0.000)} &  \\
      Education  & \multicolumn{1}{l}{0.000*} & 0.0 & \multicolumn{1}{l}{-0.001***} & -0.5 & \multicolumn{1}{l}{0.000**} & 0.0 \\
& \multicolumn{1}{l}{(0.000)} &   & \multicolumn{1}{l}{(0.000)} &   & \multicolumn{1}{l}{(0.000)} &  \\
         Log(Income) & \multicolumn{1}{l}{0.089***} & 39.7 & \multicolumn{1}{l}{0.074***} & 37.5 & \multicolumn{1}{l}{0.051***} & 36.1 \\
     & \multicolumn{1}{l}{(0.002)} &   & \multicolumn{1}{l}{(0.001)} &   & \multicolumn{1}{l}{(0.001)} &  \\
   Place of residence    & \multicolumn{1}{l}{0.074***} & 33.0 & \multicolumn{1}{l}{0.057***} & 28.9 & \multicolumn{1}{l}{0.039***} & 27.6 \\
& \multicolumn{1}{l}{(0.002)} &   & \multicolumn{1}{l}{(0.002)} &   & \multicolumn{1}{l}{(0.001)} &  \\
      & \multicolumn{1}{l}{0.027***} & 12.0 & \multicolumn{1}{l}{0.020***} & 10.1 & \multicolumn{1}{l}{0.012***} & 8.5 \\
     & \multicolumn{1}{l}{(0.002)} &   & \multicolumn{1}{l}{(0.001)} &   & \multicolumn{1}{l}{(0.001)} &  \\
 Occupation      & \multicolumn{1}{l}{0.009***} & 4.0 & \multicolumn{1}{l}{0.008***} & 4.1 & \multicolumn{1}{l}{0.007***} & 5.0 \\
    & \multicolumn{1}{l}{(0.001)} &   & \multicolumn{1}{l}{(0.001)} &   & \multicolumn{1}{l}{(0.000)} &  \\
    State    & \multicolumn{1}{l}{0.001} & 0.4 & \multicolumn{1}{l}{0.008***} & 4.1 & \multicolumn{1}{l}{0.005***} & 3.5 \\
   & \multicolumn{1}{l}{(0.002)} &   & \multicolumn{1}{l}{(0.001)} &   & \multicolumn{1}{l}{(0.001)} &  \\
    Total explained       & \multicolumn{1}{l}{} & 83.0 & \multicolumn{1}{l}{} & 76.5 & \multicolumn{1}{l}{} & 74.4 \\
      \hline 

    \hline
    \multicolumn{7}{p{50.58em}}{Notes: Internet use rate (IUR).   ***, **, and * indicate significance at the 1, 5, and 10 percent critical level.  Survey sampling weights  are for used the estimation.}
    \end{tabular}}%
  \label{tab:addlabel}%
\end{table}%

Tables 6 presents decomposition estimates of gap in the Internet literacy rate between Others and each of the disadvantaged caste groups. For example, the gap in the Internet literacy rate between Others and OBCs would have been reduced by 74.4\% if OBCs had the same characteristics as Others. Among all the covariates in the model, caste-based differences in the educational attainment of individuals are the most important drivers of the gap in the Internet literacy rate. Results from regression decomposition in Table 7, show the specific contributions of difference in  different observed characteristics to the gap in the Internet use rate between Others and each of the disadvantaged caste groups. The educational attainment of individuals is the most significant contributor to the gap in the Internet use rate between Others and each of the disadvantaged caste groups. For example, the difference in the educational attainment of the non-disadvantaged and STs explains almost 38.97\% of the total gap in the Internet use rate between these two groups. Per-capita income also contributes a good deal to the gap in the Internet use rate by individuals belonging to Other group and each of the disadvantaged caste groups. For instance, 25.5\% of the Internet use rate gap between the non-disadvantaged and OBCs can be explained through this factor.  The differences in the age distribution of the disadvantaged caste groups are favorable to them.\\

\section{Discussion and conclusion} \label{sec:Discussion and conclusions}
Using nationally representative data from Household Social Consumption: Education survey 2017-18, this paper has examined the relative contribution of caste group based disparities in different socioeconomic variables to the digital gap between Others and each of the disadvantaged caste groups (SCs, STs and OBCs). The descriptive results from the study show that there are large gaps between Others and each of the disadvantaged caste groups in computer ownership rate, Internet access rate, computer literacy rate, Internet literacy rate and Internet use rate. Especially, the digital gap between the Others and STs and SCs is very large. For example, only around 6\%  of SC and ST individuals had a computer at home as compared to 20\% of Other individuals. Similarly, the gap in computer literacy rate between Other  individuals and STs is 20 percentage points, which is the highest gap between the Others and each of the disadvantaged caste groups.\\

The decomposition analysis presented in the above section shows that more than two-thirds of the first and second-level digital gap between Others and each of the disadvantaged caste groups are attributable to differences in their socioeconomic factors. The results from the decomposition analysis reveal that caste  based digital divide can be  mostly explained by the large and persistent inequality between Others and disadvantaged caste groups in education and income. In other words, the sizeable digital divide based on caste in contemporary India reflects the historical, socio-economic backwardness of the disadvantaged caste groups (SCs, STs and OBCs).  Despite several measures including affirmative action in education and employment, educational backwardness still persists among disadvantaged caste groups in India \citep{ borooah2005vidya,deshpande2011grammar, basant2010handbook}. Further, there is ample evidence of higher level of poverty among the SCs, STs  and OBCs as compared to Others \citep{kijima2006caste, sachar2006social}. It is this educational and income disparity between the Others and the disadvantaged caste groups that emerge as the major contributor to the digital gap. In the wake of increased use of digital technologies, caste-based digital divide in India and educational and income inequality will reinforce each other.  Hence, there is an urgent need to improve educational attainments and incomes of disadvantaged caste groups to address caste-based digital divide in India.\footnote{Notably, the Appendix shows similar results for religion-based digital divide. We see that education followed by income explains more than 50\% of the total digital gap between Muslims and the non-disadvantaged group (See Appendix). }\\

The analysis of the digital divide in this paper has some limitations. First, this study only deals with the first and second-level of the digital divide. We measure the second-level digital divide based on computer and Internet skills as a binary variable. A composite index of different computer and Internet skills will better capture the second-level digital divide. Further, our analysis does not take into consideration the quality of Internet connections. Our data does not permit us to analyse gaps in the effective use of the Internet by household members in terms of gaining tangible benefits. The second limitation of our study is that our decomposition analysis cannot be interpreted as a causal analysis. Constrained by data set, the study uses limited covariates. The model does not include variables such as unobserved behavioural characteristics of individuals in using computer and the Internet. Despite these limitations, this paper has provided policy relevant empirical evidence on the factors contributing to the digital divide between Others and disadvantaged caste groups in India.\\

From a policy perspective, understanding the factors that contribute to the digital divide between Others and disadvantaged caste groups is vital to take appropriate measures toward bridging the digital divide. The results indicate some possible strategies that could aid policymakers to address caste-based digital inequalities in India. Education is a crucial factor in explaining the gap in digital skills such as knowing how to use computer and the Internet. Therefore, it is essential to improve the educational attainment of disadvantaged caste groups in India. This could be achieved by primarily focusing on younger cohorts of individuals from the disadvantaged caste group by decreasing their school dropout rate and providing more incentives to them for continuing their education. Typically, a disadvantaged caste group is more likely to be poor and less educated, and hence are less likely to have a computer at home. The economically weaker sections in developing countries may not be able to afford a computer, given the high cost of computers. One possible way to mitigate this problem is to provide subsidy to bring down the cost of computers and the Internet connection for the poor households. Some of the state governments in India have distributed free laptops to students who have graduated higher secondary schooling.\footnote{In the year 2020-21 the state government of Tamil Nadu plans to distribute 500,000 laptops to school and college going students. See https://www.thehindubusinessline.com/news/tamil-nadu-government-to-distribute-free-laptops-to-5-lakh-students-in-2020-21/article31106316.ece } Such subsidization programmes are a step in the right direction as they not only help the students, but also provide access to computer to other household members. Another way to make computer affordable to the poor households is to develop low-cost computers for the poor in developing countries. Alternatively, providing access to free computer and the Internet through community centers will help to bridge the caste-based digital divide in India . The second level digital divide can be bridged through digital literacy programmes \cite{van2014digital}.\footnote{ To address the digital skill gap in the context of developing countries a recent paper suggests two kinds of intervention -- one, directly providing digital skills to those who lack them through digital literacy programmes or, in the short run, adopting digital technologies that minimise requirement of digital skills \cite{james2019confronting}.}  The new policy of the government of Kerala to provide free high speed Internet to below poverty line households will work towards shrinking the digital gap.\footnote{In Kerala the state government initiated a programme to  provide free high speed Internet connection to over 20 lakh Below Poverty Line households see https://www.thehindu.com/news/national/kerala/govt-clearance-for-k-fon-project/article29901484.ece} Digital inequalities in India are rooted in the educational and income inequalities. Government policies that are cognizant of this fact will be closer to achieving digital equality in India

\clearpage
\singlespacing
\setlength\bibsep{0pt}
\bibliographystyle{apacite}
\bibliography{Digital}
\clearpage
\newpage 
\textbf{Appendix }

For the analysis in the Appendix, the Muslim social group includes all people who have reported their religion as Islam, irrespective of their caste. To ensure that Muslims are compared to the non-SC, ST and OBC population in other religions, the paper first divides the population into Muslims and the rest. Next, it separates the SC, ST and OBC from the non-Muslim population. For the analysis done in the Appendix, the Muslims are compared to this residual category which we refer to as the non-disadvantaged (non-SC, ST, OBC and Muslims)

\begin{table}[htbp]
  \centering
  \caption{Summary statistics}
   \resizebox{\textwidth}{!}{\begin{tabular}{p{4.215em}rrrrrrrrr}
   
    \hline
   & \multicolumn{1}{c}{(1)ND} & \multicolumn{1}{c}{(2)ST} & \multicolumn{1}{c}{(3)SC} & \multicolumn{1}{c}{(4)OBC} & \multicolumn{1}{c}{(5)Muslim} & \multicolumn{1}{c}{(1)-(2)} & \multicolumn{1}{c}{(1)-(3)} & \multicolumn{1}{c}{(1)-(4)} & \multicolumn{1}{c}{(1)-(5)} \\
       \hline

   Variables   & \multicolumn{1}{c}{Mean} & \multicolumn{1}{c}{Mean} & \multicolumn{1}{c}{Mean} & \multicolumn{1}{c}{Mean} & \multicolumn{1}{c}{Mean} & \multicolumn{1}{c}{t-test} & \multicolumn{1}{c}{t-test} & \multicolumn{1}{c}{t-test} & \multicolumn{1}{c}{t-test} \\
     \hline
    COR    & \multicolumn{1}{c}{0.241} & \multicolumn{1}{c}{0.054} & \multicolumn{1}{c}{0.066} & \multicolumn{1}{c}{0.097} & \multicolumn{1}{c}{0.082} & \multicolumn{1}{c}{0.187***} & \multicolumn{1}{c}{0.175***} & \multicolumn{1}{c}{0.144***} & \multicolumn{1}{c}{0.159***} \\
      & \multicolumn{1}{c}{[0.001]} & \multicolumn{1}{c}{[0.001]} & \multicolumn{1}{c}{[0.001]} & \multicolumn{1}{c}{[0.001]} & \multicolumn{1}{c}{[0.001]} &   &   &   &  \\
    IAR & \multicolumn{1}{c}{0.458} & \multicolumn{1}{c}{0.139} & \multicolumn{1}{c}{0.181} & \multicolumn{1}{c}{0.245} & \multicolumn{1}{c}{0.23} & \multicolumn{1}{c}{0.319***} & \multicolumn{1}{c}{0.277***} & \multicolumn{1}{c}{0.213***} & \multicolumn{1}{c}{0.228***} \\
      & \multicolumn{1}{c}{[0.002]} & \multicolumn{1}{c}{[0.002]} & \multicolumn{1}{c}{[0.002]} & \multicolumn{1}{c}{[0.001]} & \multicolumn{1}{c}{[0.002]} &   &   &   &  \\
    CLR & \multicolumn{1}{c}{0.357} & \multicolumn{1}{c}{0.11} & \multicolumn{1}{c}{0.135} & \multicolumn{1}{c}{0.196} & \multicolumn{1}{c}{0.143} & \multicolumn{1}{c}{0.247***} & \multicolumn{1}{c}{0.222***} & \multicolumn{1}{c}{0.161***} & \multicolumn{1}{c}{0.215***} \\
      & \multicolumn{1}{c}{[0.002]} & \multicolumn{1}{c}{[0.001]} & \multicolumn{1}{c}{[0.001]} & \multicolumn{1}{c}{[0.001]} & \multicolumn{1}{c}{[0.002]} &   &   &   &  \\
    ILR & \multicolumn{1}{c}{0.428} & \multicolumn{1}{c}{0.143} & \multicolumn{1}{c}{0.176} & \multicolumn{1}{c}{0.244} & \multicolumn{1}{c}{0.202} & \multicolumn{1}{c}{0.285***} & \multicolumn{1}{c}{0.252***} & \multicolumn{1}{c}{0.184***} & \multicolumn{1}{c}{0.226***} \\
      & \multicolumn{1}{c}{[0.002]} & \multicolumn{1}{c}{[0.002]} & \multicolumn{1}{c}{[0.002]} & \multicolumn{1}{c}{[0.001]} & \multicolumn{1}{c}{[0.002]} &   &   &   &  \\
    IUR & \multicolumn{1}{c}{0.391} & \multicolumn{1}{c}{0.121} & \multicolumn{1}{c}{0.15} & \multicolumn{1}{c}{0.211} & \multicolumn{1}{c}{0.179} & \multicolumn{1}{c}{0.270***} & \multicolumn{1}{c}{0.242***} & \multicolumn{1}{c}{0.180***} & \multicolumn{1}{c}{0.212***} \\
      & \multicolumn{1}{c}{[0.002]} & \multicolumn{1}{c}{[0.002]} & \multicolumn{1}{c}{[0.001]} & \multicolumn{1}{c}{[0.001]} & \multicolumn{1}{c}{[0.002]} &   &   &   &  \\
    Age & \multicolumn{1}{c}{35.13} & \multicolumn{1}{c}{33.501} & \multicolumn{1}{c}{33.542} & \multicolumn{1}{c}{34.323} & \multicolumn{1}{c}{32.501} & \multicolumn{1}{c}{1.629***} & \multicolumn{1}{c}{1.588***} & \multicolumn{1}{c}{0.807***} & \multicolumn{1}{c}{2.629***} \\
      & \multicolumn{1}{c}{[0.042]} & \multicolumn{1}{c}{[0.056]} & \multicolumn{1}{c}{[0.051]} & \multicolumn{1}{c}{[0.036]} & \multicolumn{1}{c}{[0.055]} &   &   &   &  \\
    Urban  & \multicolumn{1}{c}{0.46} & \multicolumn{1}{c}{0.123} & \multicolumn{1}{c}{0.225} & \multicolumn{1}{c}{0.281} & \multicolumn{1}{c}{0.387} & \multicolumn{1}{c}{0.337***} & \multicolumn{1}{c}{0.236***} & \multicolumn{1}{c}{0.180***} & \multicolumn{1}{c}{0.073***} \\
      & \multicolumn{1}{c}{[0.002]} & \multicolumn{1}{c}{[0.002]} & \multicolumn{1}{c}{[0.002]} & \multicolumn{1}{c}{[0.001]} & \multicolumn{1}{c}{[0.002]} &   &   &   &  \\
    Male  & \multicolumn{1}{c}{0.516} & \multicolumn{1}{c}{0.507} & \multicolumn{1}{c}{0.516} & \multicolumn{1}{c}{0.509} & \multicolumn{1}{c}{0.509} & \multicolumn{1}{c}{0.009***} & \multicolumn{1}{c}{0.000} & \multicolumn{1}{c}{0.006***} & \multicolumn{1}{c}{0.007**} \\
      & \multicolumn{1}{c}{[0.002]} & \multicolumn{1}{c}{[0.002]} & \multicolumn{1}{c}{[0.002]} & \multicolumn{1}{c}{[0.001]} & \multicolumn{1}{c}{[0.002]} &   &   &   &  \\
    MPCE  & \multicolumn{1}{c}{7.860} & \multicolumn{1}{c}{7.250} & \multicolumn{1}{c}{7.396} & \multicolumn{1}{c}{7.516} & \multicolumn{1}{c}{7.496} & \multicolumn{1}{l}{0.609***} & \multicolumn{1}{l}{0.464***} & \multicolumn{1}{l}{0.344***} & \multicolumn{1}{l}{0.364***} \\
      & \multicolumn{1}{c}{[0.002]} & \multicolumn{1}{c}{[0.002]} & \multicolumn{1}{c}{[0.002]} & \multicolumn{1}{c}{[0.002]} & \multicolumn{1}{c}{[0.002]} & \multicolumn{1}{l}{} & \multicolumn{1}{l}{} & \multicolumn{1}{l}{} & \multicolumn{1}{l}{} \\
    Education  & \multicolumn{1}{c}{10.268} & \multicolumn{1}{c}{6.363} & \multicolumn{1}{c}{7.028} & \multicolumn{1}{c}{8.087} & \multicolumn{1}{c}{6.934} & \multicolumn{1}{c}{3.905***} & \multicolumn{1}{c}{3.240***} & \multicolumn{1}{c}{2.181***} & \multicolumn{1}{c}{3.334***} \\
      & \multicolumn{1}{c}{[0.016]} & \multicolumn{1}{c}{[0.023]} & \multicolumn{1}{c}{[0.022]} & \multicolumn{1}{c}{[0.015]} & \multicolumn{1}{c}{[0.023]} &   &   &   &  \\
    SEA & \multicolumn{1}{c}{0.294} & \multicolumn{1}{c}{0.395} & \multicolumn{1}{c}{0.225} & \multicolumn{1}{c}{0.357} & \multicolumn{1}{c}{0.174} & \multicolumn{1}{c}{-0.101***} & \multicolumn{1}{c}{0.069***} & \multicolumn{1}{c}{-0.063***} & \multicolumn{1}{c}{0.120***} \\
      & \multicolumn{1}{c}{[0.002]} & \multicolumn{1}{c}{[0.002]} & \multicolumn{1}{c}{[0.002]} & \multicolumn{1}{c}{[0.001]} & \multicolumn{1}{c}{[0.002]} &   &   &   &  \\
    SENA  & \multicolumn{1}{c}{0.276} & \multicolumn{1}{c}{0.12} & \multicolumn{1}{c}{0.161} & \multicolumn{1}{c}{0.214} & \multicolumn{1}{c}{0.354} & \multicolumn{1}{c}{0.156***} & \multicolumn{1}{c}{0.115***} & \multicolumn{1}{c}{0.062***} & \multicolumn{1}{c}{-0.079***} \\
      & \multicolumn{1}{c}{[0.002]} & \multicolumn{1}{c}{[0.001]} & \multicolumn{1}{c}{[0.002]} & \multicolumn{1}{c}{[0.001]} & \multicolumn{1}{c}{[0.002]} &   &   &   &  \\
    RSW & \multicolumn{1}{c}{0.275} & \multicolumn{1}{c}{0.094} & \multicolumn{1}{c}{0.155} & \multicolumn{1}{c}{0.164} & \multicolumn{1}{c}{0.148} & \multicolumn{1}{c}{0.181***} & \multicolumn{1}{c}{0.120***} & \multicolumn{1}{c}{0.111***} & \multicolumn{1}{c}{0.127***} \\
      & \multicolumn{1}{c}{[0.002]} & \multicolumn{1}{c}{[0.001]} & \multicolumn{1}{c}{[0.002]} & \multicolumn{1}{c}{[0.001]} & \multicolumn{1}{c}{[0.002]} &   &   &   &  \\
    CWA & \multicolumn{1}{c}{0.042} & \multicolumn{1}{c}{0.185} & \multicolumn{1}{c}{0.199} & \multicolumn{1}{c}{0.097} & \multicolumn{1}{c}{0.091} & \multicolumn{1}{c}{-0.142***} & \multicolumn{1}{c}{-0.157***} & \multicolumn{1}{c}{-0.055***} & \multicolumn{1}{c}{-0.049***} \\
      & \multicolumn{1}{c}{[0.001]} & \multicolumn{1}{c}{[0.002]} & \multicolumn{1}{c}{[0.002]} & \multicolumn{1}{c}{[0.001]} & \multicolumn{1}{c}{[0.001]} &   &   &   &  \\
    CWNA & \multicolumn{1}{c}{0.055} & \multicolumn{1}{c}{0.163} & \multicolumn{1}{c}{0.22} & \multicolumn{1}{c}{0.126} & \multicolumn{1}{c}{0.18} & \multicolumn{1}{c}{-0.108***} & \multicolumn{1}{c}{-0.165***} & \multicolumn{1}{c}{-0.071***} & \multicolumn{1}{c}{-0.125***} \\
      & \multicolumn{1}{c}{[0.001]} & \multicolumn{1}{c}{[0.002]} & \multicolumn{1}{c}{[0.002]} & \multicolumn{1}{c}{[0.001]} & \multicolumn{1}{c}{[0.002]} &   &   &   &  \\
    OW & \multicolumn{1}{c}{0.039} & \multicolumn{1}{c}{0.024} & \multicolumn{1}{c}{0.019} & \multicolumn{1}{c}{0.024} & \multicolumn{1}{c}{0.027} & \multicolumn{1}{c}{0.015***} & \multicolumn{1}{c}{0.020***} & \multicolumn{1}{c}{0.015***} & \multicolumn{1}{c}{0.012***} \\
      & \multicolumn{1}{c}{[0.001]} & \multicolumn{1}{c}{[0.001]} & \multicolumn{1}{c}{[0.001]} & \multicolumn{1}{c}{[0.000]} & \multicolumn{1}{c}{[0.001]} &   &   &   &  \\
    SB1 & \multicolumn{1}{c}{0.253} & \multicolumn{1}{c}{0.254} & \multicolumn{1}{c}{0.25} & \multicolumn{1}{c}{0.246} & \multicolumn{1}{c}{0.258} & \multicolumn{1}{c}{-0.001} & \multicolumn{1}{c}{0.002} & \multicolumn{1}{c}{0.007***} & \multicolumn{1}{c}{-0.005*} \\
      & \multicolumn{1}{c}{[0.001]} & \multicolumn{1}{c}{[0.002]} & \multicolumn{1}{c}{[0.002]} & \multicolumn{1}{c}{[0.001]} & \multicolumn{1}{c}{[0.002]} &   &   &   &  \\
    SB2 & \multicolumn{1}{c}{0.252} & \multicolumn{1}{c}{0.24} & \multicolumn{1}{c}{0.242} & \multicolumn{1}{c}{0.252} & \multicolumn{1}{c}{0.248} & \multicolumn{1}{c}{0.012***} & \multicolumn{1}{c}{0.010***} & \multicolumn{1}{c}{0.000} & \multicolumn{1}{c}{0.004*} \\
      & \multicolumn{1}{c}{[0.001]} & \multicolumn{1}{c}{[0.002]} & \multicolumn{1}{c}{[0.002]} & \multicolumn{1}{c}{[0.001]} & \multicolumn{1}{c}{[0.002]} &   &   &   &  \\
    SB3 & \multicolumn{1}{c}{0.252} & \multicolumn{1}{c}{0.268} & \multicolumn{1}{c}{0.253} & \multicolumn{1}{c}{0.248} & \multicolumn{1}{c}{0.255} & \multicolumn{1}{c}{-0.016***} & \multicolumn{1}{c}{-0.001} & \multicolumn{1}{c}{0.004*} & \multicolumn{1}{c}{-0.003} \\
      & \multicolumn{1}{c}{[0.001]} & \multicolumn{1}{c}{[0.002]} & \multicolumn{1}{c}{[0.002]} & \multicolumn{1}{c}{[0.001]} & \multicolumn{1}{c}{[0.002]} &   &   &   &  \\
    SB4 & \multicolumn{1}{c}{0.243} & \multicolumn{1}{c}{0.239} & \multicolumn{1}{c}{0.255} & \multicolumn{1}{c}{0.254} & \multicolumn{1}{c}{0.239} & \multicolumn{1}{c}{0.005*} & \multicolumn{1}{c}{-0.011***} & \multicolumn{1}{c}{-0.011***} & \multicolumn{1}{c}{0.004*} \\
      & \multicolumn{1}{c}{[0.001]} & \multicolumn{1}{c}{[0.002]} & \multicolumn{1}{c}{[0.002]} & \multicolumn{1}{c}{[0.001]} & \multicolumn{1}{c}{[0.002]} &   &   &   &  \\
  N & \multicolumn{1}{c}{85258} & \multicolumn{1}{c}{47340} & \multicolumn{1}{c}{57256} & \multicolumn{1}{c}{117600} & \multicolumn{1}{c}{48416} &   &   &   &  \\
   
        \hline
    \multicolumn{10}{p{50.15em}}{Notes: Computer ownership rate (COR); Internet access rate (IAR); Computer literacy rate (CLR); Internet literacy rate (ILR); Internet use rate (IUR); Log per-capita income(MPCE); Self-employed in agriculture (SEA); Self-employed in non-agriculture (SENA);Regular salaried Workers (RSW); Casual worker in agriculture (CWA); Casual worker in non-agriculture (CWNA); Other work(OW); Observation(N); Non-disadvantaged(ND); Sub-round(SB). Standard errors are in parentheses. ***, **, and * indicate significance at the 1, 5, and 10 percent critical level. Sampling weights of the survey used for the calculations.} \\
    
    \end{tabular}}%
  \label{tab:addlabel}%
\end{table}%
 
\begin{table}[htbp]
  \centering
  \caption{Marginal effects from probit regressions on the probability of having computer at home, internet access at home, computer literacy, Internet literacy and Internet use. }
   \resizebox{\textwidth}{!}{\begin{tabular}{llllll}
  \hline

      \hline 
      Variables& 	CO &	AIF & 	CL& IL &IU \\
   \hline 
  
    Age & 0.001*** & 0.002*** & -0.003*** & -0.006*** & -0.005*** \\
      & (0.000) & (0.000) & (0.000) & (0.000) & (0.000) \\
    Gender: reference = Male &   \\
    Female & 0.011*** & 0.026*** & -0.022*** & -0.069*** & -0.067*** \\
      & (0.001) & (0.003) & (0.001) & (0.002) & (0.002) \\
  Place of residence: reference  = Rural& \\
    Urban & 0.022*** & 0.061*** & 0.016*** & 0.036*** & 0.033*** \\
      & (0.002) & (0.004) & (0.001) & (0.003) & (0.002) \\
    Social group: reference = ND&   \\
    ST & -0.037*** & -0.117*** & -0.025*** & -0.058*** & -0.049*** \\
      & (0.003) & (0.006) & (0.002) & (0.003) & (0.003) \\
    SC & -0.043*** & -0.088*** & -0.024*** & -0.054*** & -0.047*** \\
      & (0.003) & (0.005) & (0.002) & (0.003) & (0.003) \\
    OBC & -0.042*** & -0.070*** & -0.017*** & -0.038*** & -0.034*** \\
      & (0.002) & (0.004) & (0.002) & (0.003) & (0.002) \\
    Muslim & -0.046*** & -0.078*** & -0.025*** & -0.046*** & -0.038*** \\
      & (0.003) & (0.005) & (0.002) & (0.003) & (0.003) \\
    Education & 0.008*** & 0.020*** & 0.017*** & 0.032*** & 0.026*** \\
      & (0.000) & (0.000) & (0.000) & (0.000) & (0.000) \\
    Log per-capita income & 0.072*** & 0.121*** & 0.031*** & 0.067*** & 0.062*** \\
      & (0.002) & (0.003) & (0.002) & (0.002) & (0.002) \\
  Occupation : reference = SEA&   \\
    SENA & 0.041*** & 0.068*** & 0.011*** & 0.030*** & 0.032*** \\
      & (0.003) & (0.005) & (0.001) & (0.003) & (0.003) \\
    CWA & -0.025*** & -0.070*** & -0.006*** & -0.017*** & -0.015*** \\
      & (0.002) & (0.005) & (0.001) & (0.003) & (0.002) \\
    CWNA & -0.025*** & -0.058*** & -0.009*** & -0.008** & -0.006** \\
      & (0.002) & (0.005) & (0.001) & (0.003) & (0.002) \\
    RSW & 0.051*** & 0.096*** & 0.027*** & 0.056*** & 0.058*** \\
      & (0.003) & (0.005) & (0.002) & (0.003) & (0.003) \\
    OW & 0.024*** & 0.027** & 0.034*** & 0.058*** & 0.047*** \\
      & (0.005) & (0.009) & (0.004) & (0.007) & (0.006) \\
   Month  of interview: reference = SB1&  \\
    SB2 & -0.011*** & 0.004 & -0.007*** & -0.004 & 0.002 \\
      & (0.002) & (0.004) & (0.001) & (0.002) & (0.002) \\
    SB3 & -0.016*** & 0.032*** & -0.006*** & 0.005* & 0.010*** \\
      & (0.002) & (0.004) & (0.001) & (0.002) & (0.002) \\
    SB4 & -0.016*** & 0.081*** & -0.007*** & 0.015*** & 0.026*** \\
      & (0.002) & (0.004) & (0.001) & (0.002) & (0.002) \\
    State controls  & Yes & Yes & Yes & Yes & Yes \\
    \hline
    \multicolumn{6}{p{40.15em}}{Notes: Computer ownership (CO); 	Access to Internet facility(AIF);  Computer literacy (CL); Internet literacy  (IL); Internet use  (IU); Self-employed in agriculture (SEA); Self-employed in non-agriculture (SENA);Regular salaried workers  (RSW); Casual worker in agriculture (CWA);  Casual worker in non-agriculture (CWNA); Other work(OW); Sub-round(SB); Non-disadvantaged(ND). Robust standard errors are in parentheses. ***, **, and * indicate significance at the 1, 5, and 10 percent critical level. Sampling weights of the survey used for the estimation.}\\

    \end{tabular}}
  \label{ tab: addlabel}%
\end{table}

\begin{table}[htbp]
  \centering
  \caption{Nonlinear decompositions of social group disparities in computer ownership rate}
     \resizebox{\textwidth}{!}{\begin{tabular}{lllllllll}
 \hline

    \hline
    Variables & ND-ST & \% explained & ND-SC & \% explained & ND-OBC & \% explained & ND-Muslims & \% explained \\
     \hline
    
    Gap in COR & 0.187 &   & 0.175 &   & 0.144 &   & 0.159 &  \\
    Age  & 0.003*** & 1.6 & 0.003*** & 1.7 & 0.002*** & 1.4 & 0.004*** & 2.5 \\
      & (0.000) &   & (0.000) &   & (0.000) &   & (0.000) &  \\
    Gender  & 0.000 & 0.0 & 0.000** & 0.0 & 0.000 & 0.0 & -0.000* & 0.0 \\
      & (0.000) &   & (0.000) &   & (0.000) &   & (0.000) &  \\
    Education & 0.042*** & 23 & 0.039*** & 22.9 & 0.027*** & 19.4 & 0.046*** & 28.3 \\
      & (0.001) &   & (0.001) &   & (0.001) &   & (0.001) &  \\
    Log per-capita income & 0.082*** & 43.4 & 0.074*** & 42.9 & 0.055*** & 37.5 & 0.063*** & 39.6 \\
      & (0.003) &   & (0.002) &   & (0.001) &   & (0.002) &  \\
    Place of residence  & 0.033*** & 17.7 & 0.021*** & 12 & 0.015*** & 10.4 & 0.007*** & 3.8 \\
      & (0.002) &   & (0.002) &   & (0.001) &   & (0.001) &  \\
    Occupation & 0.011*** & 5.9 & 0.012*** & 7.4 & 0.010*** & 6.9 & 0.005*** & 3.1 \\
      & (0.001) &   & (0.001) &   & (0.001) &   & (0.001) &  \\
    Sub-round & -0.000 & 0.0 & 0.000 & 0.0 & 0.000*** & 0.0 & -0.000 & 0.0 \\
      & (0.000) &   & (0.000) &   & (0.000) &   & (0.000) &  \\
    State  & -0.002 & -1.6 & -0.017*** & -10.3 & -0.015*** & -9.7 & -0.009*** & -5 \\
      & (0.002) &   & (0.001) &   & (0.002) &   & (0.001) &  \\
    Total explained &   & 90 &   & 76.6 &   & 65.9 &   & 72.4 \\
     \hline
  
    \multicolumn{9}{p{50.58em}}{Notes: Computer ownership rate(COR); Non-disadvantaged(ND).   ***, **, and *indicate significance at the 1, 5, and 10 percent critical level. Survey sampling weights are used for the estimation}
    \end{tabular}}%
  \label{tab:addlabel}%
\end{table}%

\begin{table}[htbp]
    \centering
  \caption{Nonlinear decompositions of social group disparities in Internet access rate}
  \resizebox{\textwidth}{!}{\begin{tabular}{lllllllll} 
 \hline

    \hline
    Variables & ND-ST & \% explained & ND-SC & \% explained & ND-OBC & \% explained & ND-Muslims & \% explained \\
     \hline
   
    Gap IAR & 0.319 &   & 0.277 &   & 0.213 &   & 0.228 &  \\
    Age  & 0.003*** & 0.9 & 0.003*** & 1.4 & 0.002*** & 0.8 & 0.005*** & 2.2 \\
      & (0.000) &   & (0.000) &   & (0.000) &   & (0.001) &  \\
    Gender  & -0.000 & 0.0 & 0.000 & 0.0 & -0.000 & 0.0 & -0.000** & 0.0 \\
      & (0.000) &   & (0.000) &   & (0.000) &   & (0.000) &  \\
    Education & 0.074*** & 23.5 & 0.065*** & 23.1 & 0.044*** & 18.0 & 0.068*** & 30.0 \\
      & (0.002) &   & (0.002) &   & (0.001) &   & (0.002) &  \\
    Log per-capita income  & 0.094*** & 29.5 & 0.066*** & 23.8 & 0.053*** & 21.6 & 0.059*** & 25.7 \\
      & (0.004) &   & (0.002) &   & (0.002) &   & (0.002) &  \\
    Place of residence  & 0.042*** & 13.2 & 0.029*** & 10.5 & 0.020*** & 8.2 & 0.009*** & 3.9 \\
      & (0.003) &   & (0.002) &   & (0.001) &   & (0.001) &  \\
    Occupation & 0.011*** & 3.4 & 0.017*** & 6.1 & 0.010*** & 4.1 & 0.010*** & 4.3 \\
      & (0.001) &   & (0.001) &   & (0.001) &   & (0.001) &  \\
    Sub-round & -0.000*** & 0.0 & -0.001*** & -0.4 & -0.001*** & -0.4 & 0.000*** & 0.0 \\
      & (0.000) &   & (0.000) &   & (0.000) &   & (0.000) &  \\
    State  & 0.017*** & 5.0 & 0.018*** & 6.5 & 0.020*** & 8.6 & 0.017*** & 7.0 \\
      & (0.003) &   & (0.002) &   & (0.002) &   & (0.002) &  \\
    Total explained  &   & 75.5 &   & 71.1 &   & 60.8 &   & 73.1 \\
    Observations & 132596 &   & 142509 &   & 202856 &   & 133671 &  \\
    \hline
   
    \multicolumn{9}{p{55.58em}}{Notes: Internet access  rate(IAR); Non-disadvantaged(ND). ***, **, and * indicate significance at the 1, 5, and 10 percent critical level.Survey sampling weights are used for the estimation}
    \end{tabular}}%
  \label{tab:addlabel}%
\end{table}%

\begin{table}[htbp]
  \centering

  \caption{Nonlinear decompositions of social group disparities in computer literacy rate}
   
    \resizebox{\textwidth}{!}{\begin{tabular}{lllllllll}
     \hline
 
    \hline
    Variables & ND-ST & \% explained & ND-SC & \% explained & ND-OBC & \% explained & ND-Muslims & \% explained \\
     \hline
    
    Gap in CLR & 0.247&&	0.222&&	0.161&&	0.215\\

     Age  & -0.016*** & -6.5 & -0.015*** & -7.7 & -0.009*** & -5.0 & -0.023*** & -10.2 \\
    & (0.000) &   & (0.000) &   & (0.000) &   & (0.001) &  \\
       Gender& 0.000 & 0.0 & -0.000*** & 0.0 & -0.000 & 0.0 & 0.001*** & 0.5 \\
     & (0.000) &   & (0.000) &   & (0.000) &   & (0.000) &  \\
      Education & 0.116*** & 47.0 & 0.097*** & 44.1 & 0.068*** & 41.6 & 0.112*** & 52.2 \\
    & (0.002) &   & (0.002) &   & (0.001) &   & (0.002) &  \\
       Log per-capita income & 0.070*** & 28.3 & 0.056*** & 25.2 & 0.043*** & 26.1 & 0.047*** & 21.9 \\
    & (0.003) &   & (0.002) &   & (0.001) &   & (0.002) &  \\
       Place of residence & 0.027*** & 10.9 & 0.020*** & 9.5 & 0.014*** & 8.7 & 0.008*** & 3.7 \\
    & (0.002) &   & (0.001) &   & (0.001) &   & (0.001) &  \\
     Occupation   & 0.014*** & 5.7 & 0.012*** & 5.4 & 0.009*** & 5.6 & 0.009*** & 4.2 \\
   & (0.001) &   & (0.001) &   & (0.001) &   & (0.001) &  \\
       Sub-round  & -0.000* & 0.0 & 0.000 & 0.0 & 0.000*** & 0.0 & -0.000 & 0.0 \\
  & (0.000) &   & (0.000) &   & (0.000) &   & (0.000) &  \\
        State   & 0.000 & 0.0 & 0.003* & 1.4 & -0.001 & 0.0 & 0.005*** & 2.3 \\
 & (0.002) &   & (0.001) &   & (0.001) &   & (0.001) &  \\
     Total explained  &   & 85.4 &   & 77.9 &   & 77.1 &   & 74.5 \\
    Observations & 132595 &   & 142509 &   & 202856 &   & 133671 &  \\
    \hline
    
    \multicolumn{9}{p{55.58em}}{Notes: Computer literacy  rate(CLR); Non-disadvantaged(ND).  ***, **, and * indicate significance at the 1, 5, and 10 percent critical level. Survey sampling weights are used for the estimation}
    \end{tabular}}%
  \label{tab:addlabel}%
\end{table}%

\begin{table}[htbp]
  \centering
  
 \caption{Nonlinear decompositions of social group disparities in Internet literacy rate}
  \resizebox{\textwidth}{!}{\begin{tabular}{lllllllll}
    \hline

    \hline
    Variables & ND-ST & \% explained & ND-SC & \% explained & ND-OBC & \% explained & ND-Muslims & \% explained \\
     \hline
 
    Gap in ILR& 0.285&&	0.252&&	0.184&&	0.226\\

   Age & -0.017*** & -5.6 & -0.016*** & -6.8 & -0.008*** & -4.4 & -0.023*** & -9.3 \\
      & (0.000) &   & (0.000) &   & (0.000) &   & (0.000) &  \\
      &   &   &   &   &   &   &   &  \\
    Gender  & 0.001*** & 0.4 & -0.000 & 0.0 & 0.000* & 0.0 & 0.002*** & 0.9 \\
      & (0.000) &   & (0.000) &   & (0.000) &   & (0.000) &  \\
    Education & 0.132*** & 46.7 & 0.110*** & 44.5 & 0.075*** & 40.8 & 0.119*** & 52.2 \\
      & (0.002) &   & (0.002) &   & (0.001) &   & (0.002) &  \\
    Log per-capita income  & 0.075*** & 26.4 & 0.057*** & 22.3 & 0.044*** & 23.9 & 0.048*** & 20.8 \\
      & (0.003) &   & (0.002) &   & (0.001) &   & (0.002) &  \\
    Place of residence  & 0.028*** & 9.8 & 0.021*** & 8.3 & 0.014*** & 7.6 & 0.009*** & 4.0 \\
      & (0.002) &   & (0.001) &   & (0.001) &   & (0.001) &  \\
    Occupation & 0.012*** & 4.2 & 0.010*** & 4.0 & 0.008*** & 4.4 & 0.005*** & 2.2 \\
      & (0.001) &   & (0.001) &   & (0.001) &   & (0.001) &  \\
    Sub-round & -0.000*** & 0.0 & -0.000* & 0.0 & -0.000*** & 0.0 & 0.000** & 0.0 \\
      & (0.000) &   & (0.000) &   & (0.000) &   & (0.000) &  \\
    State  & -0.000 & 0.0 & 0.010*** & 4.5 & 0.005*** & 3.1 & 0.009*** & 4.2 \\
      & (0.002) &   & (0.001) &   & (0.001) &   & (0.001) &  \\
    Total explained  &   & 81.9 &   & 76.8 &   & 75.5 &   & 75.0 \\
    Observations & 132595 &   & 142509 &   & 202856 &   & 133671 &  \\
    
       \hline
    \multicolumn{9}{p{50.58em}}{Notes. Internet literacy rate (ILR); Non-disadvantaged(ND).  ***, **, and * indicate significance at the 1, 5, and 10 percent critical level. Survey sampling weights are for used the estimation.}
    \end{tabular}}%
  \label{tab:addlabel}%
\end{table}%

\begin{table}[htbp]
  \centering
\caption{Nonlinear decompositions of social group disparities in Internet use rate}
    \resizebox{\textwidth}{!}{\begin{tabular}{lllllllll}
 
    \hline
     Variables & ND-ST & \% explained & ND-SC & \% explained & ND-OBC & \% explained & ND-Muslims & \% explained \\

    \hline
    Gap in IUR & 0.27 &   & 0.242 &   & 0.18 &   & 0.212 &  \\
    Age  & -0.016*** & -5.9 & -0.016*** & -6.6 & -0.009*** & -5.0 & -0.021*** & -11.7 \\
      & (0.000) &   & (0.000) &   & (0.000) &   & (0.000) &  \\
    Gender  & 0.001*** & 0.4 & -0.000** & 0.0 & 0.000*** & 0.6 & 0.002*** & 1.1 \\
      & (0.000) &   & (0.000) &   & (0.000) &   & (0.000) &  \\
    Education & 0.112*** & 41.5 & 0.095*** & 39.3 & 0.065*** & 36.2 & 0.104*** & 58.0 \\
      & (0.002) &   & (0.002) &   & (0.001) &   & (0.002) &  \\
    Log per-capita income  & 0.081*** & 30.0 & 0.062*** & 25.2 & 0.046*** & 25.6 & 0.052*** & 29.0 \\
      & (0.003) &   & (0.002) &   & (0.001) &   & (0.002) &  \\
    Place of residence  & 0.031*** & 11.5 & 0.023*** & 9.5 & 0.015*** & 8.3 & 0.009*** & 5.0 \\
      & (0.002) &   & (0.002) &   & (0.001) &   & (0.001) &  \\
    Occupation & 0.011*** & 4.1 & 0.010*** & 4.1 & 0.010*** & 5.6 & 0.005*** & 2.8 \\
      & (0.001) &   & (0.001) &   & (0.001) &   & (0.001) &  \\
    Sub-round & -0.000* & 0.0 & -0.000* & 0.0 & -0.000*** & 0.0 & 0.000*** & 0.0 \\
      & (0.000) &   & (0.000) &   & (0.000) &   & (0.000) &  \\
    State  & 0.002 & 0.7 & 0.011*** & 4.6 & 0.008*** & 4.5 & 0.009*** & 5.0 \\
      & (0.002) &   & (0.001) &   & (0.002) &   & (0.001) &  \\
    Total explained  & 132595 &   & 142509 &   & 202856 &   & 133671 &  \\
    
       \hline
    \multicolumn{9}{p{50.58em}}{Notes. Internet use rate (IUR); Non-disadvantaged(ND).  ***, **, and * indicate significance at the 1, 5, and 10 percent critical level.Bootstrap standard errors in parentheses.  Survey sampling weights  are for used the estimation.}
    \end{tabular}}%
  \label{tab:addlabel}%
\end{table}%

\end{document}